\documentclass[lettersize,journal]{IEEEtran}
\usepackage{amsmath,amsfonts}
\usepackage{array}
\usepackage[caption=false,font=normalsize,labelfont=sf,textfont=sf]{subfig}
\usepackage{textcomp}
\usepackage{stfloats}
\usepackage{url}
\usepackage{verbatim}
\usepackage{graphicx}
\usepackage{cite}
\usepackage{ulem}
\usepackage{xcolor}
\DeclareMathOperator*{\spn}{span}

\begin{document}

\title{Road abnormality detection using piezoresistive force sensors and adaptive signal models}

\author{Tam\'as D\'ozsa,
\and
J\'anos Rad\'o*\thanks{*Tam\'as D\'ozsa and J\'anos Rad\'o contributed equally to this work and should both be considered first authors of the paper.},
\and
J\'anos Volk,
\and
\'Ad\'am Kisari,
\and
Alexandros Soumelidis,
\and
P\'eter Kov\'acs

\thanks{
Tam\'as D\'ozsa (dotuaai@inf.elte.hu) and P\'eter Kov\'acs (kovika@inf.elte.hu) are researchers at the Department of Numerical Analysis, Faculty of Informatics E\"otv\"os Lor\'and University (Budapest, Hungary). J\'anos Rad\'o (rado.janos@ek-cer.hu) and J\'anos Volk (volk.janos@ek-cer.hu) are reaserchers at the Nanosensors Laborartory of the Centre for Energy Research (Budapest, Hungary). Tam\'as D\'ozsa (dozsa.tamas@sztaki.hu), \'Ad\'am Kisari (kisari.adam@sztaki.hu) and Alexandros Soumelidis (soumelidis@sztaki.mta.hu) are researchers at the Systems and Control Laboratory of the Institute of Computer Science and Control (Budapest, Hungary).
}

}

\markboth{Journal of IEEE Transactions on Instrumentation and Measurement }%
{Shell \MakeLowercase{\textit{et al.}}: A Sample Article Using IEEEtran.cls for IEEE Journals}

\IEEEpubid{}

\maketitle

\begin{abstract}
Intelligent tires can be employed for a wide array of applications ranging from tire pressure monitoring  to analyzing tire/road interactions, wheel loading as well as tread wear monitoring. In this paper we develop a measurement system for intelligent tires equipped with a 3-dimensional piezoresistive force sensor. The output of the sensor is segmented into tire revolution cycles, which are then represented by a transformation relying on adaptive Hermite functions. The underlying idea behind this step is to extract relevant features which capture tire dynamics. Then we evaluate the proposed measurement system in a potential vehicle application, that is, abnormal road surface detection. We deal with the corresponding binary classification problem by developing both low-complexity analytical and data-driven machine learning algorithms, which are tested on real-world measurement data. Our experiments showed that the proposed methods are able to detect abnormalities on the road surface with a mean accuracy of over $97 \%$.
\end{abstract}


\section{Introduction}

\IEEEPARstart{V}{ehicular} automation has been making considerable progress in the last decade setting new requirements for sensing technologies. In this respect, environmental recognition is a key source of information to replace human control. The driving conditions are observed by fusing the output of various sensors mounted on the car, such as cameras, radar and LIDAR. The concept of smart tires is another step towards more accurate environmental sensing. For instance, tire pressure monitoring systems are already well-known and infiltrated the mass market. In addition to tire pressure, modern sensors provide information about tire-road interactions which can be used in vehicle applications, such as structural health monitoring of the tires, terrain classification, and optimization of active safety systems (e.g., ABS, ESP). 

State of the art technologies for intelligent tire systems include optical sensors, accelerometers, strain sensors, acoustic wave sensors, and polyvinylidene fluoride (PVDF) sensors~\cite{apollo, smart_tire_review1}. Each of these approaches has advantages and disadvantages in monitoring the deformation of the tire. For instance, even though accelerometers are compact, energy efficient, and inexpensive sensors, they can track only one specific point on the tire, and the measured signal is contaminated by the noise generated from the road surface as well as rotational, vibrational and gravitational accelerations~\cite{smart_tire_review2}. In contrast, strain sensors are not affected by the rotational speed of the tire, and they proved to be a better approach to estimate wheel forces \cite{erdogan_estimation_2011}. Wheel forces can also be estimated 
by using optical sensors~\cite{tiredeformation2, tiredeformation1}. In this measurement setup, a proper alignment between the detector and the light source is necessary. Hence, the sensor system must be recalibrated after any misalignment caused by abrupt driving control, such as severe breaking~\cite{smart_tire_review2}.

In this study we introduce a wireless light weight measurement system, which is compact in size (i.e., it does not influence the balance and stability of the tire), robust against abrupt tire deformations, vibrations caused by the engine, and gravitational accelerations. The proposed force sensor is based on piezoresistive technology and estimates the direction and magnitude of the mechanical forces acting on the tire by measuring the subsequent changes in the electrical resistance of silicone components. Although metal based strain gauges, as inexpensive and reliable sensors, are widely used to detect local strain on rigid and flexible materials, their gauge factor is about ($g\approx $ 1.5-2) \cite{Bao2000} and are not sensitive to normal force component. In contrast, our three-dimensional piezoresistive force sensor has a significantly higher gauge factor ($g\approx $ 140) and is sensitive to both lateral ($F_x$, $F_y$) and vertical ($F_z$) force components providing more information on the local deformation. 

In piezoresistive 3D force sensors strain sensitive resistors are placed either on four suspended microbeams \cite{adam_cmos_2008, beccai_design_2005, liu_micro-force_2017, omiya_micropillar_2015, tibrewala_simulation_2008, quan_design_2015} or on a full membrane \cite{alcheikh_characterization_2013, molnar_sensitivity_2012}. In order to transform the acting shear force into membrane deformation, either a perpendicular stylus is mounted on the Si chip \cite{liu_micro-force_2017, tibrewala_development_2009, quan_design_2015} or a monolithic Si microrod is fabricated by bulk micromachining \cite{alcheikh_characterization_2013, beccai_design_2005, molnar_sensitivity_2012}. The reported 3D sensors have been used to solve engineering problems across a wide array of fields including medical~\cite{Rado2018, AboueiMehrizi2014, Karthikeyan2019}, industrial~\cite{Friedrich2021, Dilibal2021, 9447752} robotic, and various civil engineering \cite{9358187, Kultongkham2021} applications which rely heavily on information provided by such sensors. In this work, for the first time, a 3D force sensor is integrated into a car tire to collect and analyse information from its cyclic deformation. In order to enhance the mechanical robustness of the force sensor, the double-side processed Si was anodically bonded to a glass substrate and a novel flexible packaging technique was applied.

As a case study, the problem of road surface abnormality detection is discussed, which is a well studied and important environmental recognition problem in vehicle control. The benefits of accurate road surface abnormality detection are twofold. Firstly, combined with global position information, detected abnormalities can be reported to a centralized database greatly reducing maintenance costs and increasing the efficiency of traffic control \cite{DQGH, CLTW, ZKCG, WSDM}. On the other hand, one can estimate the overall road quality from the frequency of detected abnormalities and apply changes to the dynamical characteristics of the vehicle (e.g., speed, acceleration, suspension)~\cite{BMGS}. In the second part of the paper, we show that the proposed wheel sensor can be effectively used to detect road surface abnormalities (e.g., potholes, bumps). To this end, we develop analytical and data-driven models which rely on the data obtained from the proposed wheel sensor.

The rest of this paper is organized as follows. In Section~\ref{sec:sens_desc}, we describe the characteristics of the proposed force sensor, the technical details on tire integration, and the interpretation of the measurement data. Section~\ref{sec:exper} provides information on our test vehicle and the installation of our measurement system. In section~\ref{sec:preproc}, we discuss the preprocessing steps applied to the output of the sensor, followed by a low-dimensional Hermite-representation of the data in section~\ref{sec:herm_model}. In section~\ref{sec:classifiers}, we introduce analytic and data-driven indirect measurement schemes \cite{SAO} for road surface abnormality detection. Then, we specify the classification methods along with their hyperparameter selection in section~\ref{sec:classSpecs}. The discussion of the experiments and the results can be found in section~\ref{sec:tests}. Finally, section~\ref{sec:conclusion} is a summary of conclusions and future plans.

\section{Sensor description}
\label{sec:sens_desc}

\begin{figure}[!t]
    \centering
    \includegraphics[width=0.99\linewidth, trim={0, 0, 0, 0}, clip]{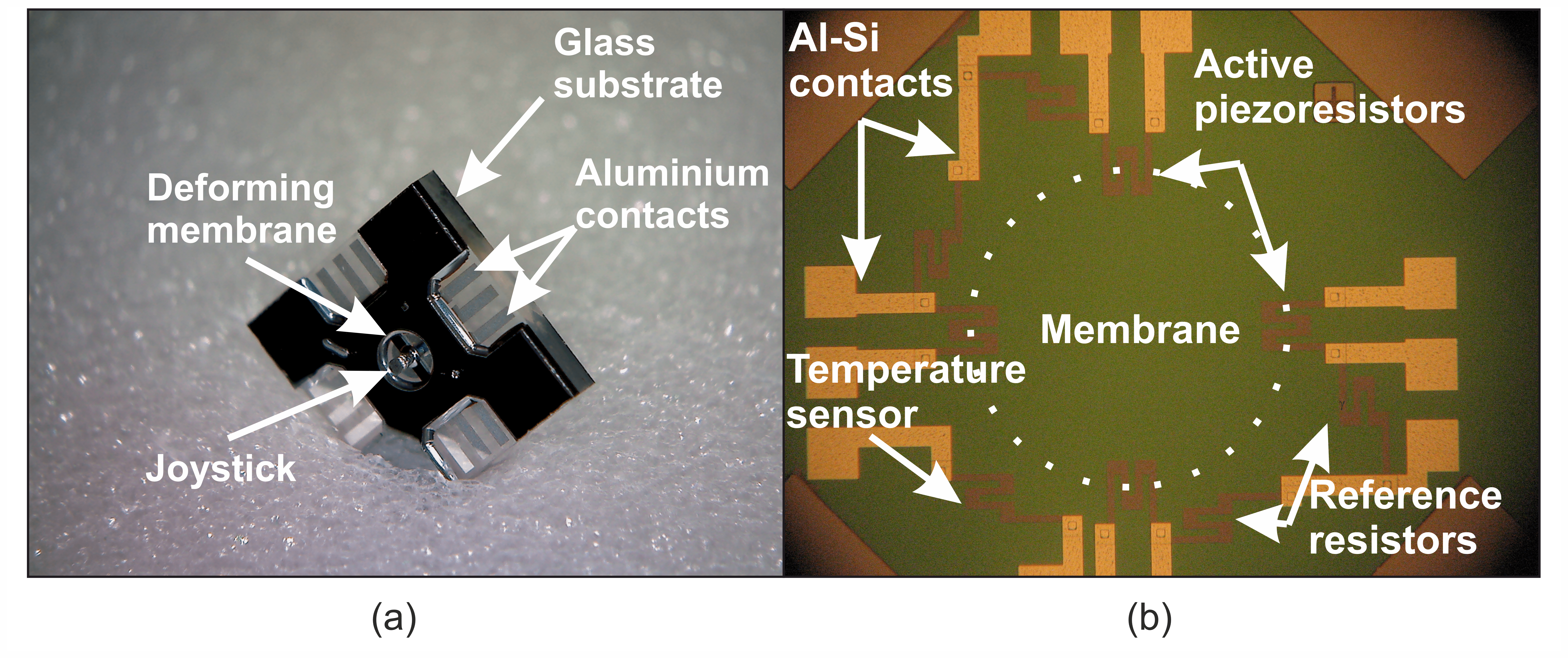}
    \caption{Silicon based 3D force sensor from the front (a) and the back (b). The anodically bonded glass substrate increases the mechanical stability of the sensor, provides a cavity underneath the deforming membrane and ensures wire contacts required for assembly.}
    \label{fig:force_sensor}
\end{figure}

The 3D force sensor (Figure \ref{fig:force_sensor}) was fabricated using conventional MEMS technology steps. A double-side-polished, (100) Si-on-Insulator (SOI) wafer, having a 50 ${\mu}$m thick n-type doped device layer, was processed on both sides and bonded anodically to a boron glass wafer. On the device side of the SOI wafer eight Si resistors were formed by boron implantation. Four of them were placed on the strained membrane along $<110>$ crystallographic directions, while 4 further ones on bulk Si to provide references. The four sets of static and sensing piezoresistors are then connected into half Wheatstone-bridges to provide maximum out-of-balance voltages upon deformation of the membrane. After that, by means of a two-step deep reactive ion etching (DRIE) process, a circular Si membrane and a concentrically positioned cylindrical joystick was released from the base side of the SOI wafer. The diameter of the emerging joystick and that of the full membrane disk is 250 ${\mu}$m and 500 ${\mu}$m, respectively. The role of the bonded glass substrate is threefold, it renders mechanical stability for the chip, provides excess for the wire bondings, and limits the maximal deformation for the membrane by forming a cavity beneath it.

The relationship between the change in bridge voltages and the force components is linear, that is described by the following equations:

\vspace{-2mm}
\begin{equation}
    \label{eq:force_sens}
    \begin{split}
        & F_x = \frac{1}{v_0 \alpha_{ls} \pi_{44}} (\Delta V_{right} - \Delta V_{left}), \\
        & F_y = \frac{1}{v_0 \alpha_{ls} \pi_{44}} (\Delta V_{top} - \Delta V_{bottom}), \\
        & F_z = \frac{1}{v_0 \alpha_{ln} \pi_{44}} \times \\
        & (\frac{\Delta V_{right} + \Delta V_{left} + \Delta V_{top} + \Delta V_{bottom}}{2}), 
    \end{split}
\end{equation}

\noindent where $F_x, F_y$ and $F_z$ are the tangential ($X$ and $Y$) and the normal ($Z$) force components, $V_0$ and $\Delta V_{positions}$ denote the common and the measured voltages at each bridge, $\pi_{44}$ is the dominant piezoresistive coefficient, $\alpha_{ln}$ and $\alpha_{ls}$ are the linear normal and shear coefficients in the given geometric arrangement. It is important to emphasize, that the forces in (\ref{eq:force_sens}) refer to the loads acting on the sensor itself, the relationship between the forces acting on the tires of the vehicle and the output signals of the embedded sensor is more complex and interrelated.

\begin{figure}[!t]
    \centering
    \includegraphics[width=0.99\linewidth, trim={0, 0, 0, 0}, clip]{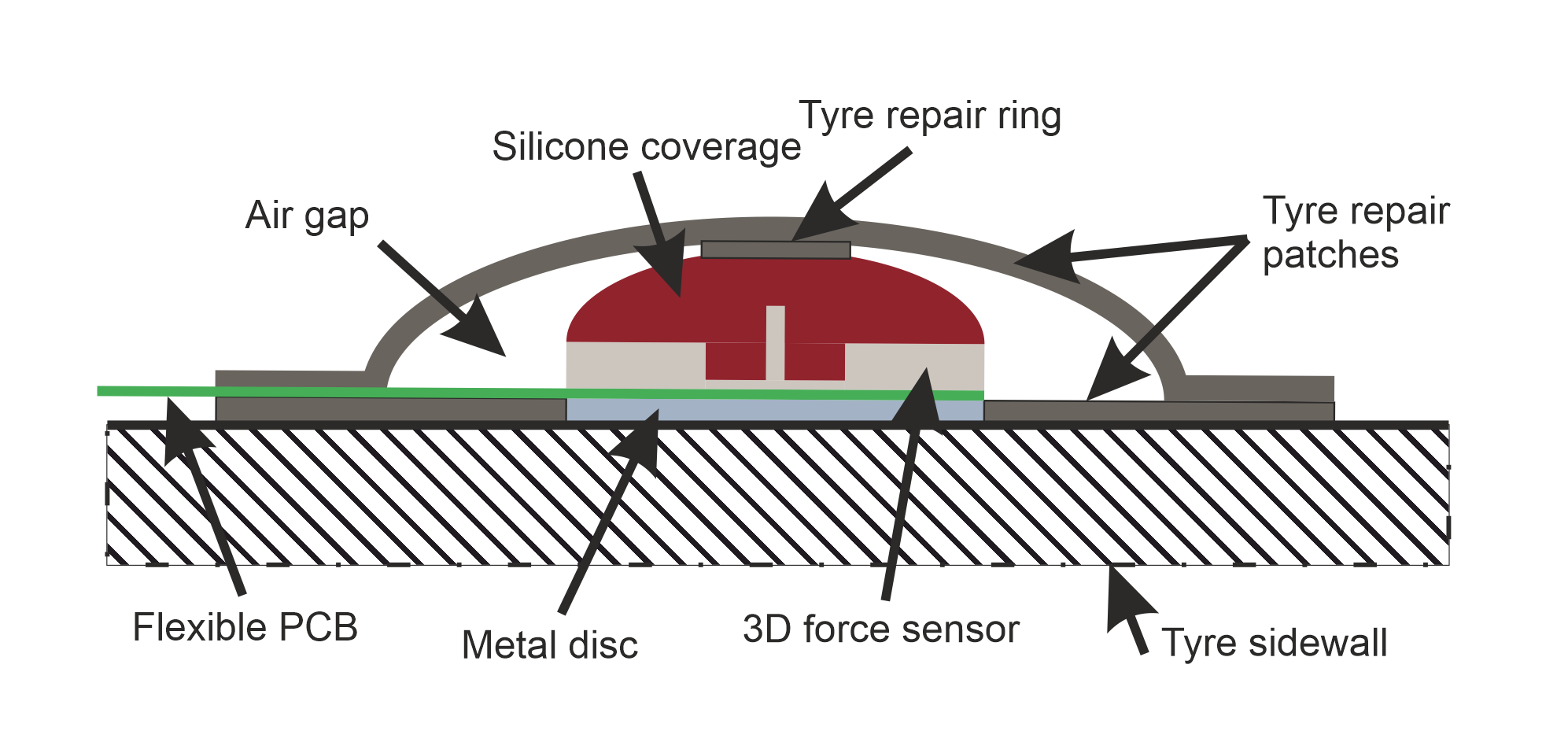}
    \caption{Schematic of the embedding process. Due to the air gap, the sidewall deformation of the tire can be transmitted to the joystick through the tire repair ring and the flexible silicone coverage.}
    \label{fig:embedding_proc}
\end{figure}

As for measuring tire deformation, there are two crucial issues, the position and the method of the embedding. To meet these requirements, we have worked out a patented multi-step process \cite{RadoPatent} that can be followed in Figure \ref{fig:embedding_proc}.

After mounting and wire-bonding the force sensor on a flexible printed circuit board (in short PCB), a hemispherical polymer protecting coverage was moulted on the top of the sensor. In order to prevent turning the PCB out from the seat while the vehicle is running, a metal disc was glued to the back of the board. The metal disc was fixed to the inner sidewall of the tire with adhesive in the seat formed of a tire repair patch. Another tire repair patch was glued to this rubber and to the silicone coverage of the sensor, leaving an air gap between the patches. This air gap ensures that the sensor membrane deforms according to the sidewall of the tire.
In order to protect the sensor from the outer harsh environment and obtain the highest response signals, the inner sidewall was chosen as the embedding location. Experience has shown that the greatest deformation within the sidewall occurs near the tread, as a result, the sensor was glued there (Figure \ref{fig:embedded_sensor}).

\section{Vehicle integration}
\label{sec:exper}

\begin{figure}[!t]
    \centering
    \includegraphics[width=0.99\linewidth, trim={0, 0, 0, 0}, clip]{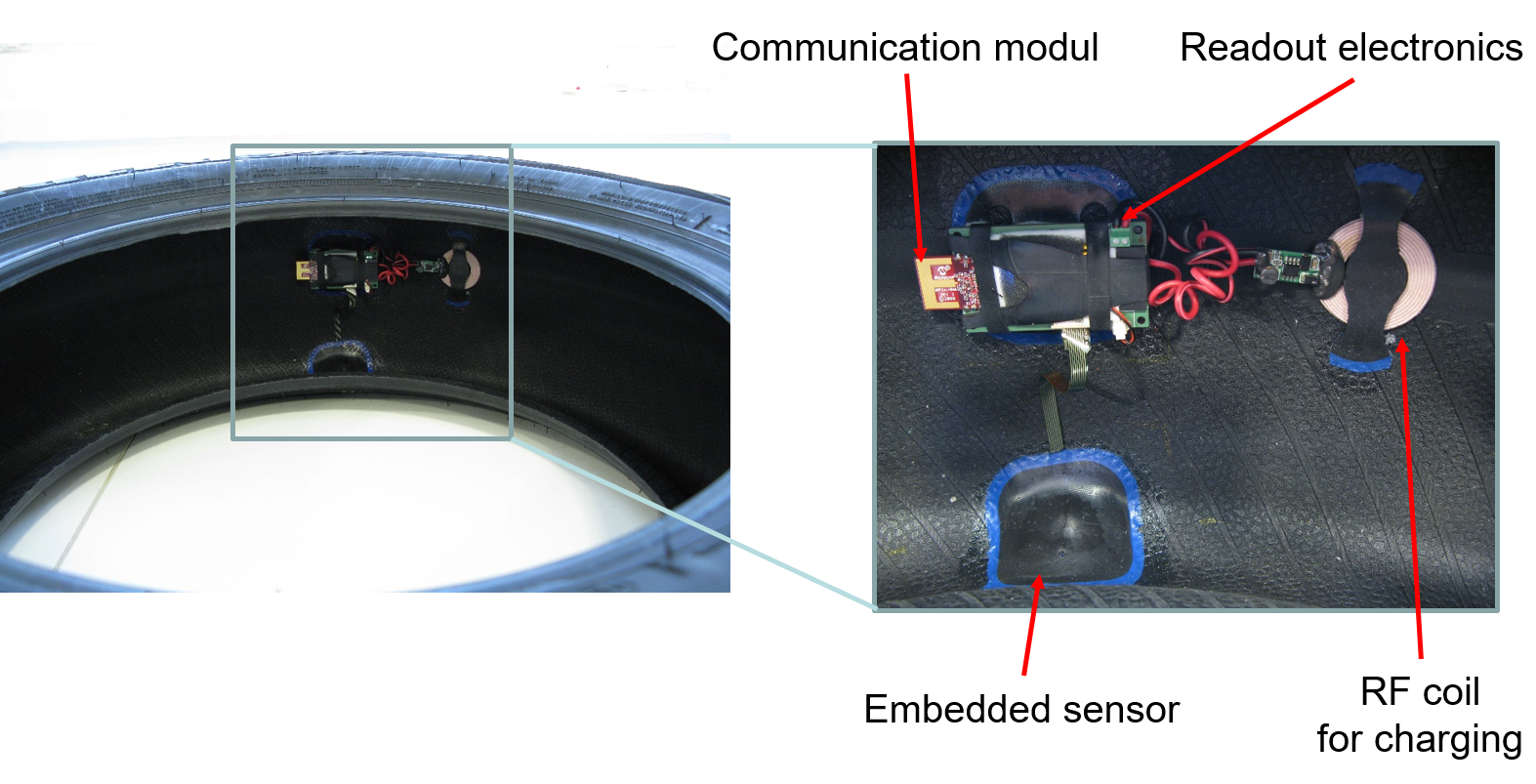}
    \caption{The complete device integrated in a tire. Tire repair patches were also used here to fix the readout electronics and the RF coil to the tire tread. To avoid wire breaks due to the relative movement between the sensor and the accessories, flexible PCB and spiral cables were applied.}
    \label{fig:embedded_sensor}
\end{figure}

To retrieve the data from the sensor built into the inner sidewall of the tire, we developed a readout electronics capable of driving the voltage bridges with 2.5V, conditioning the signals, and transmitting the data wirelessly. The readout electronics was powered by LiPo battery that can be charged from outside the tire with a commercially available RF charging coil. Both of the electronics and the RF coil were fixed to the tread as a less compliant area of the tire (Figure \ref{fig:embedded_sensor}).

\begin{figure}[!t]
    \centering
    \includegraphics[width=0.99\linewidth, trim={0, 0, 0, 0}, clip]{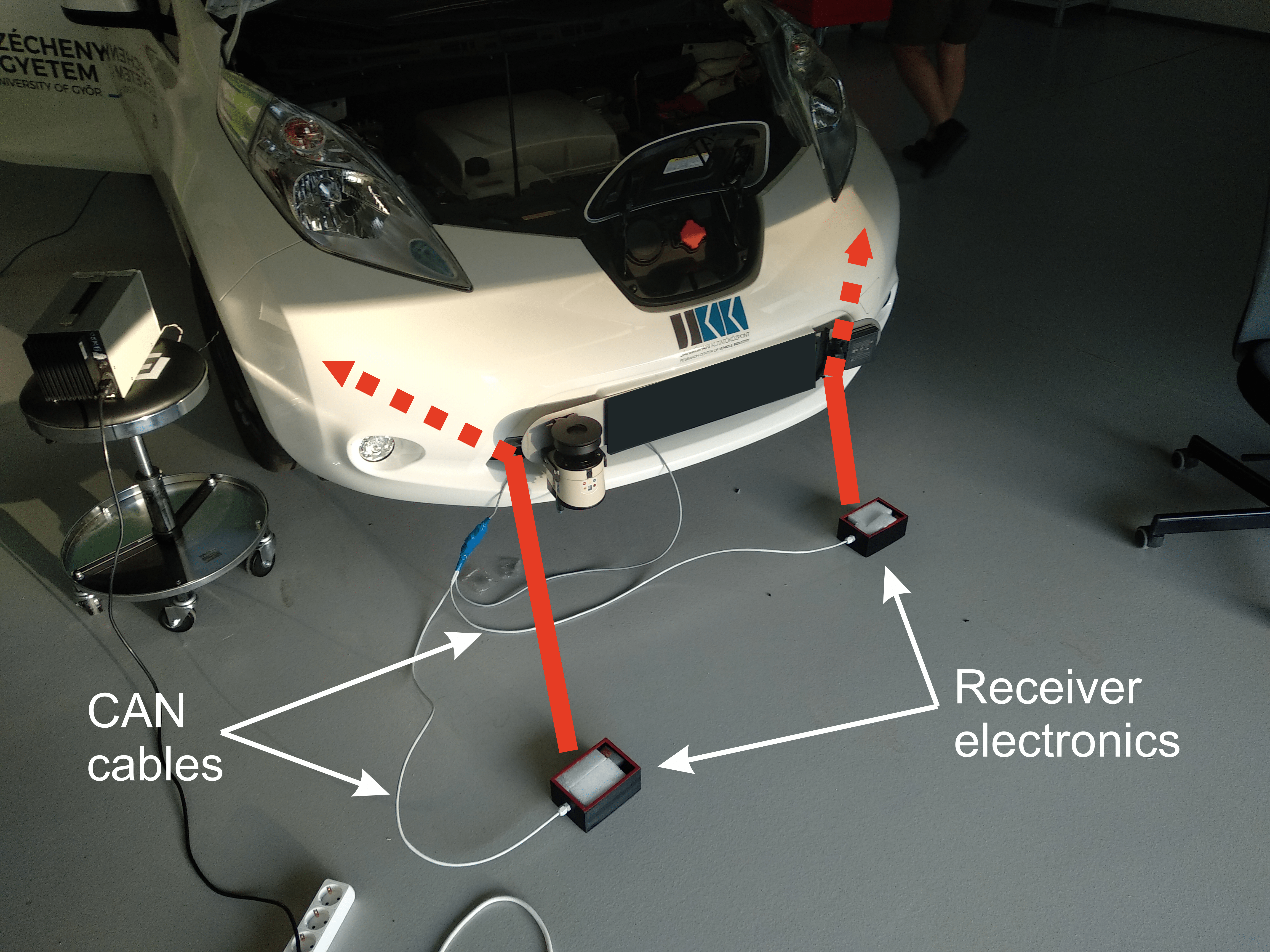}
    \caption{Off-tire receiver and data conversion electronics. The PCBs were placed in a 3D printed box to protect them from the harsh environment. The red arrows indicate the final position of the electronics.}
    \label{fig:receiver_elect}
\end{figure}

In order to carry out the real-world experiments, two test tires were mounted on a Nissan Leaf provided by Institute for Computer Science and Control. The external receiver electronics were placed in the engine bay of the car as close as possible to the tires to ensure the continuous signal transmission, while still protecting the receiver from damage (Figure \ref{fig:receiver_elect}). The CAN bus wiring was routed from the trunk to the front, between the on-board computer and the receivers, with a power supply connected near the computer in the trunk. The received signals were encoded according to our own CAN BUS protocol for local processing and data recording. The CAN communication ensures the synchronization between the tire sensor signals and the other dynamics data, such as accelerations, speed, angular momentum and GPS coordinates. These signals are partly collected from our own sensors installed on the vehicle, using our own CAN network and partly from the stock vehicle sensors that we can access.

For the final test, two different, well-separable road surfaces were chosen in Budapest. One, the so-called “normal” surface, was a newly paved, perfect asphalt pavement, while the other, “abnormal” surface, was an old, poor quality concrete road with potholes. We passed through both pavements many times and recorded the signals of the tire sensors along with other dynamic data from both our own and the vehicle CAN buses. We note that the velocity of the vehicle can influence the shape of the signals produced by the sensor (see Fig. \ref{fig:speedCompared}). In order to comply with traffic safety regulations and to ensure equipment as well as personnel safety, measurements were considered with vehicle speeds ranging from $4$ km/h to $56$ km/h.

Our test tires had a circumference of $199$ cm and the above described equipment recorded data with a sampling rate of $1000$ Hertz. In this setup, if we consider a constant vehicle velocity of $50$ km/h, a single tire revolution would be represented by $143$ data points. This is a sufficient amount of data for modeling the sensor output using the representation described in section~\ref{sec:herm_model}. In fact, similar signal models were used in \cite{KBHH} to represent the so-called QRS-complexes of ECG recordings. These signals consisted of $100$ data points and shared morphological similarities with the wheel sensor measurements.

\section{Data preprocessing}
\label{sec:preproc}

When the vehicle is in motion, the implanted force sensor produces a quasi-periodic, quasi-compact signal for each revolution of the tire. In order to use wheel sensor based signals for surface abnormality detection, we analyze the properties of each full period (corresponding to a full rotation of the tire), thus we need to segment the measurements. To achieve this, we adapted a discrete wavelet transform based segmentation algorithm \cite{JSSU} originally created for use with ECG signals. We exploited the close spatial and behavioral resemblance between ECG signals and the output of the implanted force sensor. We note that our test vehicle is currently being equipped with accurate wheel-angle measuring sensors, therefore algorithmic segmentation of the measurements may not be necessary in the future. Each segmented measurement was allocated a "normal" or an "abnormal" label using ground truth data. Our method of obtaining ground truth information is detailed in section \ref{sec:tests}.

\begin{figure}[!t]
    \centering
    \includegraphics[width=0.99\linewidth, trim={0, 0, 0, 0}, clip]{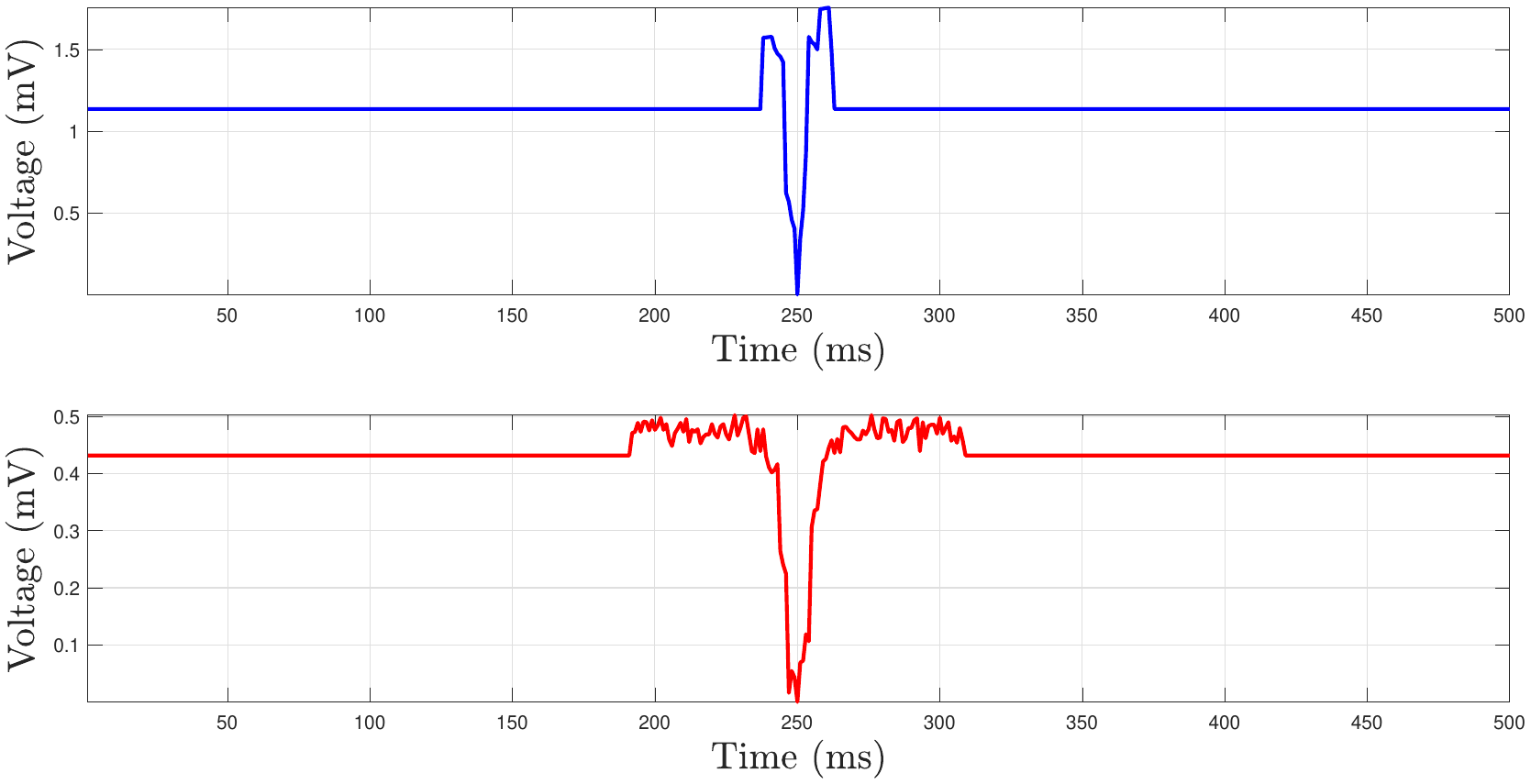}
    \caption{Preprocessed, tire implanted force sensor outputs, representing a single revolution of the wheel. \textbf{TOP:} A preprocessed signal segment labeled "normal". \textbf{BOTTOM:} A preprocessed signal segment labeled "abnormal".}
    \label{fig:norm_abnorm_periods}
\end{figure}

In order to periodize the resulting signal segments, we removed the baseline connecting the first and the last sample values from each segment. Note that the number of data points which make up a single rotation of the wheel changes with the vehicle speed. Therefore, zero padding was also applied to equalize the sampling points in each segmented period. The maximal length of a period was identified as $500$ data points, any periods longer than this (for example if the vehicle stood still for sometime) were disregarded. Periods of fewer than $500$ data points, were zero padded to match this length. Figure \ref{fig:norm_abnorm_periods} illustrates two example measurements, each representing a single revolution of the tire.

\section{Modeling sensor data using adaptive Hermite functions}
\label{sec:herm_model}

An intuitive assumption for wheel sensor based road abnormality detection is that measurements corresponding to abnormal road conditions will contain more noise than signals measured on a normal surface. This is to be expected, since on an abnormal, bumpy surface the tire of the vehicle is subjected to quickly changing mechanical forces. As a consequence, the noise level of the measurements can be a good feature to detect road surface abnormalities. For this reason, we propose a transformation to fit and remove the smooth part of the signal. In section \ref{subsec:resid}, we demonstrate that measuring the noise levels on the transformed residuals allows for more precise surface abnormality detection methods. We now proceed to describe adaptive Hermite functions as introduced in \cite{DK} and provide insight into why transformations based on them are especially useful when processing tire sensor output signals.

Hermite functions have a long history of successful applications in signal processing. The original Hermite function system was first shown to be an effective tool for modeling the so-called QRS complexes in ECG signals \cite{SLB}. Since this first application, subsequent generalizations of Hermite functions have been used to design adaptive orthogonal transformations for several biological signals \cite{JOL, SAS, DK, KBDJ, carlpeter}. These transformations formed the basis of many segmentation, illness identification and classification tasks \cite{KBJ, DBK}. In section \ref{sec:preproc} we already alluded to the similarities between the force sensor's output and several types of biological signals such as ECG measurements and action potentials. We now proceed to further exploit these similarities and describe an adaptive Hermite function based orthogonal transformation which forms the basis of our proposed road surface abnormality detection algorithms.

Let us denote the $n$-th Hermite polynomial \cite{SZ} by $h_n \ (n \in \mathbb{N})$. These polynomials are orthogonal in the weighted Lebesgue-space $L_{w}^{2}(\mathbb{R})$, where $w(x) := e^{-x^2}$. Using the Hermite polynomials we can define the so-called Hermite functions as

\begin{equation}
    \label{eq:herm_funcs}
    \Phi_k(x) = h_k(x) \cdot e^{-x^2/2}/\sqrt{\pi^{1/2} 2^k k!}, \quad (k \in \mathbb{N}),
\end{equation}

\noindent
which form an orthonormal and complete function system in $L_2(\mathbb{R})$, with respect to the inner product and norm

\begin{equation*}
    \langle f, g \rangle = \int_{-\infty}^{\infty} f(x) g(x) dx, \quad \|f\|_2 = \sqrt{\langle f, f \rangle},
\end{equation*}
where $f, g \in L^{2}(\mathbb{R})$.

\begin{figure}[!t]
    \centering
    \includegraphics[width=0.99\linewidth, trim={0, 0, 0, 0}, clip]{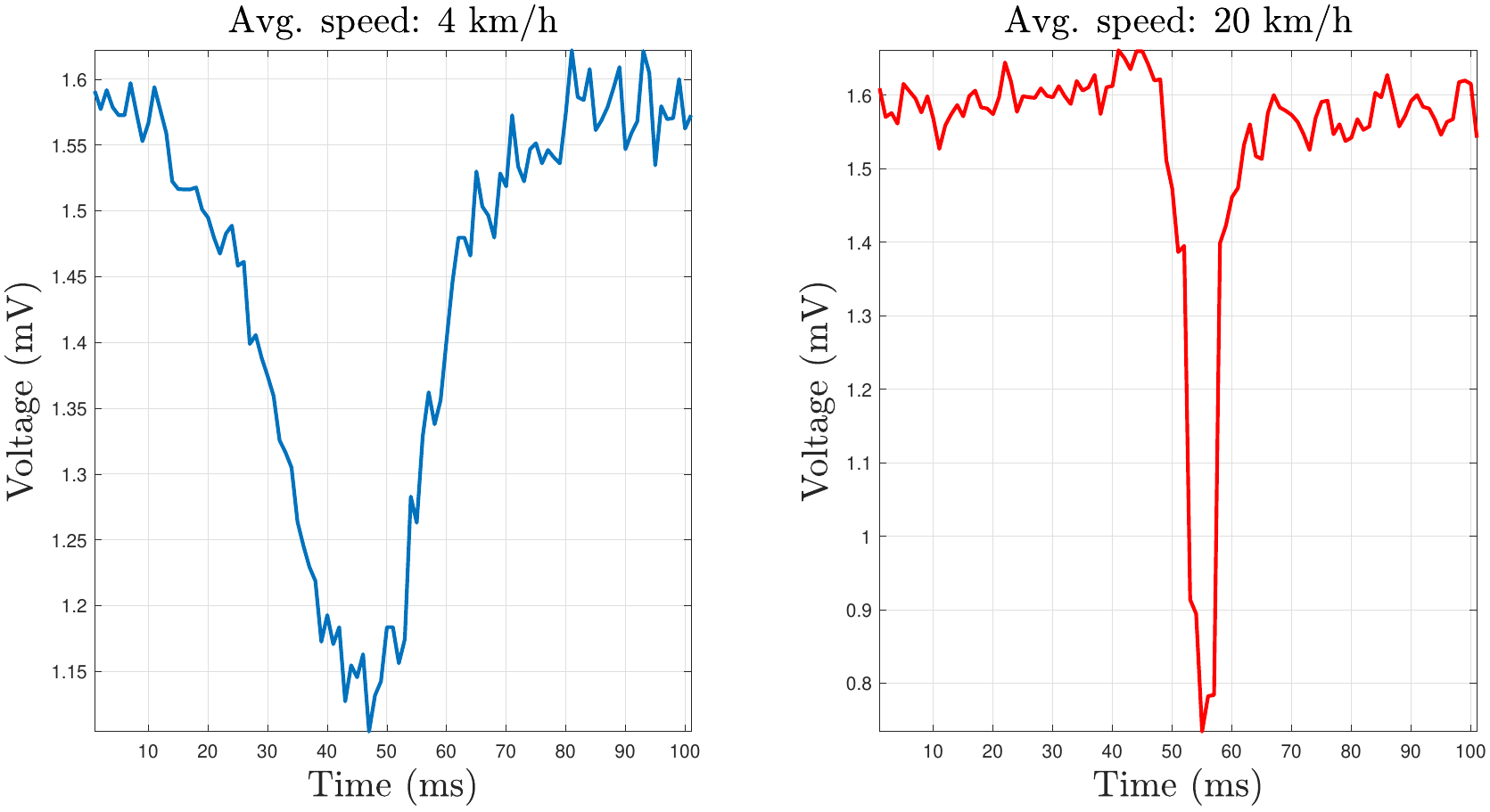}
    \caption{The width of the main waveform in a tire sensor segment and the relative location of the the peak on the time axis changes with the vehicle's speed. The two signals depicted above were sampled over equal length time intervals.  \textbf{LEFT:} Tire sensor output from a single revolution of the tire at low speeds. \textbf{RIGHT:} Tire sensor output from a single revolution of the tire at higher speeds.}
    \label{fig:speedCompared}
\end{figure}

It is well-known that the best approximation from the subspace $\spn \{ \Phi_k \ : \ 0\leq k < n\}\subset L^{2}(\mathbb{R})$ to any function $f \in L^{2}(\mathbb{R})$ can be computed by the Hermite-Fourier partial sums:
\begin{equation}
    \label{eq:smooth_apr}
    f \approx \hat{f} = \sum_{k=0}^{n-1} \langle f, \Phi_k \rangle \Phi_k \qquad (n \in \mathbb{N}).
\end{equation}
Such smooth approximations of compact or quasi-compact signals can be especially precise, because the modulus of the functions $\Phi_k(x)$ decays exponentially as $|x|$ tends to infinity. As mentioned above, many biological signal processing applications choose Hermite functions for modelling purposes, since their shapes resemble characteristic waveforms found in ECG and EEG signals (such as QRS complexes). Practically, this means that a low number of Hermite functions is enough to represent the smooth part of the signal, i.e., $n$ in \eqref{eq:smooth_apr} is expected to be small. The quasi compact property and the shape similarity with the first few Hermite functions also hold for segmented force sensor measurements. However, these are non-stationary signals caused by the dynamics of the vehicle. Indeed, the speed changes affect the width and the location of the main spike-like waveform in each rotation cycle (see e.g. Figure~\ref{fig:speedCompared}).

\begin{figure}[!t]
    \centering
    \includegraphics[width=0.99\linewidth, trim={0, 0, 0, 0}, clip]{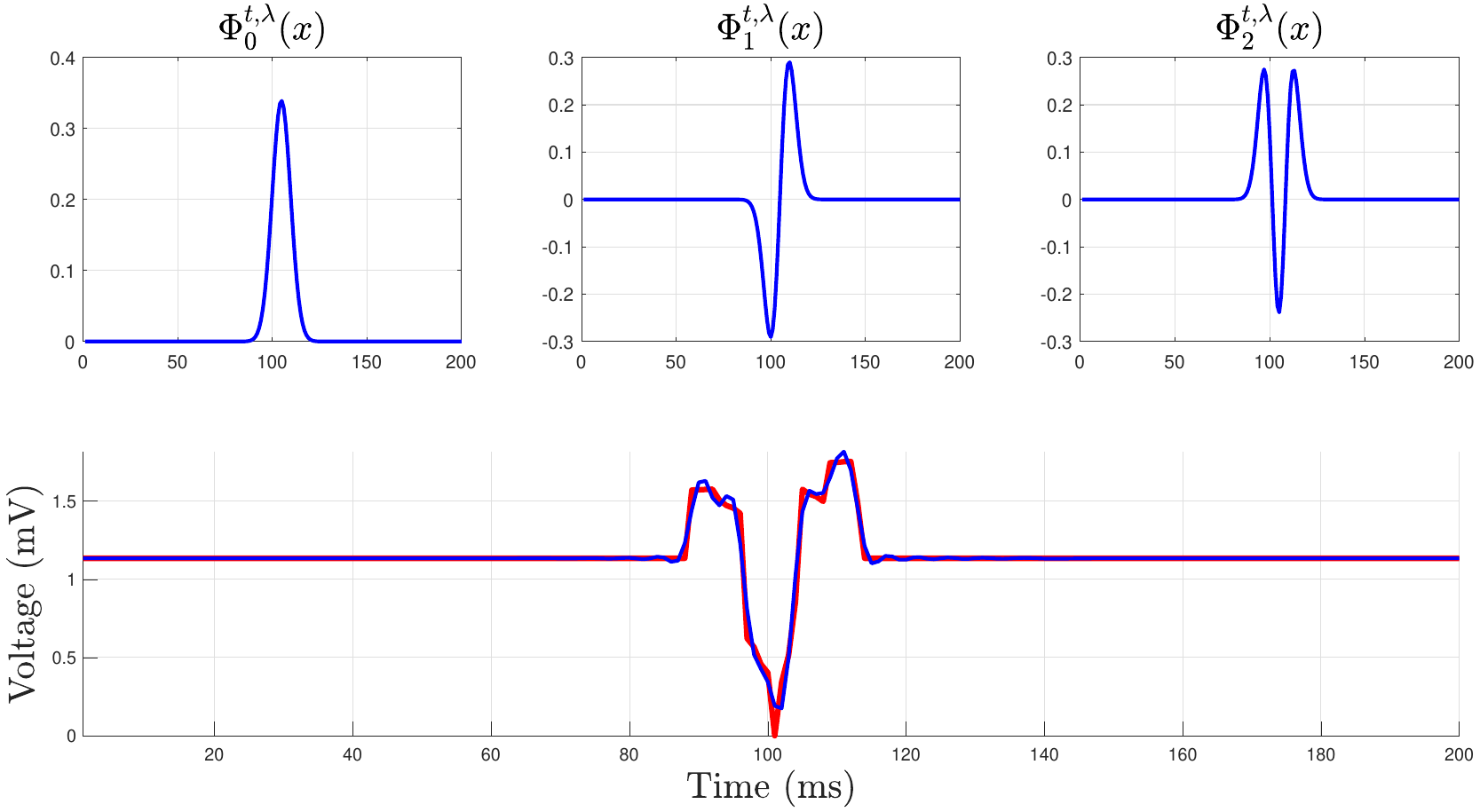}
    \caption{\textbf{TOP:} The first three adaptive Hermite functions for fixed dilation and translation parameters. \textbf{BOTTOM:} Approximation of a wheel sensor output segment (red) with a linear combination of adaptive Hermite functions (blue).}
    \label{fig:hermfuns}
\end{figure}

In order to model this phenomena, we use the affine argument transformation of the classical Hermite functions:

\begin{equation}
	\Phi_k^{\tau,\lambda}(x):= \sqrt{\lambda} \Phi_k(\lambda (x - \tau)) \quad (x,\tau\in\mathbb{R}, \lambda>0)\, \textrm{,}
	\label{eq:affHerm}
\end{equation}
where $\Phi_k^{\tau,\lambda}$'s are referred to as adaptive Hermite functions \cite{DK}. These systems also retain the complete and orthonormal properties in $L^2(\mathbb{R})$. In addition, this adaptive system inherits the properties (quasi compact behavior, shape similarities) that make Hermite functions suitable for the approximation of segmented force sensor signals. We note that adaptive Hermite functions and their derivatives as well as partial derivatives with respect to the $\lambda$ dilation and  $\tau$ translation parameters can be calculated using stable three term recurrence formulas. As described below, this will allow us to rely on gradient-based methods to identify the optimal parameters of the adaptive Hermite function system. Figure \ref{fig:hermfuns} illustrates some adaptive Hermite functions and the subsequent approximation of a wheel sensor output segment.

In applications we can represent a force sensor measurement segment by $\boldsymbol{y} \in \mathbb{R}^N, \ (N \in \mathbb{N})$. The components of $\boldsymbol{y}$ can be thought of as values of the unknown function $f \in L^2(\mathbb{R})$ describing the sensor output sampled at specific time instants. Consider the matrix $\boldsymbol{\Phi}^{\lambda, \tau} \in \mathbb{R}^{N \times n}$, whose columns are made up of discrete samplings of the first $n$ adaptive Hermite functions. Our goal is to find the best possible approximation of $\boldsymbol{y}$ in the form
\begin{equation}
    \boldsymbol{y} \approx \boldsymbol{\Phi}^{\lambda, \tau} \boldsymbol{c} = \sum_{k=0}^{n-1} c_k \cdot \boldsymbol{\varphi}^{\lambda, \tau}_k \textrm{,}
    \label{eq:linearrep}
\end{equation}
where, $\boldsymbol{\varphi}^{\lambda, \tau}_k \in \mathbb{R}^N$ stand for the discretized adaptive Hermite functions, and $c_k$ denotes the $k$th coordinate of $\boldsymbol{c}\in\mathbb{R}^n$. For any given $\lambda$ and $\tau$, the best linear least squares approximation can be calculated by $\boldsymbol{c}=\left( \boldsymbol{\Phi}^{\lambda, \tau}\right)^+\boldsymbol{y}$, where $\left( \boldsymbol{\Phi}^{\lambda, \tau}\right)^+$ denotes the Moore--Penrose pseudoinverse of $\boldsymbol{\Phi}^{\lambda, \tau}$. Thus, finding a good approximation to the sensor output results in the following optimization problem:
\begin{equation}
    \label{eq:varpro}
    \begin{split}
        \min_{\lambda > 0, \tau \in \mathbb{R}} \| \boldsymbol{y} - \left(\boldsymbol{\Phi}^{\lambda, \tau}\right) \left( \boldsymbol{\Phi}^{\lambda, \tau}\right)^+ \boldsymbol{y} \|_{2}^{2}.
    \end{split}
\end{equation}
Note that the transformed signal is the orthogonal projection of $\boldsymbol{y}$ onto the column space of $\boldsymbol{\Phi}^{\lambda, \tau}$, hence the operator
\begin{equation}
    \label{eq:proj}
     P_{\boldsymbol{\Phi}^{\lambda, \tau}}(\boldsymbol{y}) = \left(\boldsymbol{\Phi}^{\lambda, \tau}\right) \left( \boldsymbol{\Phi}^{\lambda, \tau}\right)^+ \boldsymbol{y}
\end{equation} 
is referred to as a variable projection operator. Since the partial derivatives of the adaptive Hermite functions in~\eqref{eq:affHerm} exist with respect to $\lambda$ and $\tau$, ~\eqref{eq:varpro} can be solved by using gradient-based methods \cite{GP}. In addition to the orthogonal projections one can consider the orthogonal complement of $P_{\boldsymbol{\Phi}^{\lambda, \tau}}(\boldsymbol{y})$, referred to as the residual signal:

\begin{equation}
    \label{eq:resid}
    P_{\boldsymbol{\Phi}^{\lambda, \tau}}^{\perp}(\boldsymbol{y}) = \boldsymbol{y} - P_{\boldsymbol{\Phi}^{\lambda, \tau}}(\boldsymbol{y}).
\end{equation}
This will be of particular interest to us, as in section \ref{sec:classifiers}, we develop surface abnormality detection schemes, which depend on the residual signal of each segment. We sum up the proposed adaptive orthogonal residual transformation with the following steps.

\begin{enumerate}
    \item Solve (\ref{eq:varpro}) to determine the optimal dilation $\lambda^*$ and translation $\tau^*$ parameters for the given signal segment $\boldsymbol{y}$. Note that this is different from section \ref{subsec:vpnet}, where we discuss a data driven approach to find $\lambda^*$ and $\tau^*$ by maximizing the abnormality detection accuracy for a set of signal segments.
    \item Approximate $\boldsymbol{y}$ with $P^{\lambda^*, \tau^*}_{\boldsymbol{\Phi}} (\boldsymbol{y})$ given in ~\eqref{eq:proj}.
    \item Compute the residual $P_{\boldsymbol{\Phi}^{\lambda, \tau}}^{\perp}(\boldsymbol{y})$ given in \eqref{eq:resid}.
\end{enumerate}

\section{Road surface Abnormality Recognition}
\label{sec:classifiers}

We are going to pose the problem of surface abnormality detection as a binary classification task. In order to specify and evaluate our classification schemes, we need to define a set of force sensor measurement segments $\mathcal{F}$ along with the corresponding ground truth class labels. In our case, a segment $\boldsymbol{y}\in\mathcal{F}$ can only be labelled as "normal" or "abnormal" depending on whether it was recorded on a normal road surface. Then, we can think of different classifiers (or classification models) as functions which map measurement segments onto said labels. To evaluate the proposed classifiers we will use the notion of model accuracy. We note that the model accuracy for classifiers given as

\begin{equation}
    \label{eq:acc}
    \text{Accuracy}=\frac{TP + TN}{P + N}
\end{equation}
is different from the accuracy notion usually considered in instrumentation and measurement problems \cite{SAO}. In ~\eqref{eq:acc}, the notations "TP" and "TN" refer to the number of times the classifier correctly identified a measurement segment as abnormal (TP or true positive) or as normal (TN or true negative). The notations "P" and "N" refer to the total number of abnormal and normal segments in the set to which the classifier was applied. In this section first we demonstrate the utility of the Hermite function based adaptive residual transformation~\eqref{eq:resid} through an analytic classifier in \ref{subsec:resid}. Then, we extend this idea to a data-driven classifier in \ref{subsec:vpnet} to further improve model accuracy. 

\subsection{An analytic approach}
\label{subsec:resid}

\begin{figure}[!t]
    \centering
    \includegraphics[width=0.99\linewidth, trim={0, 0, 0, 0}, clip]{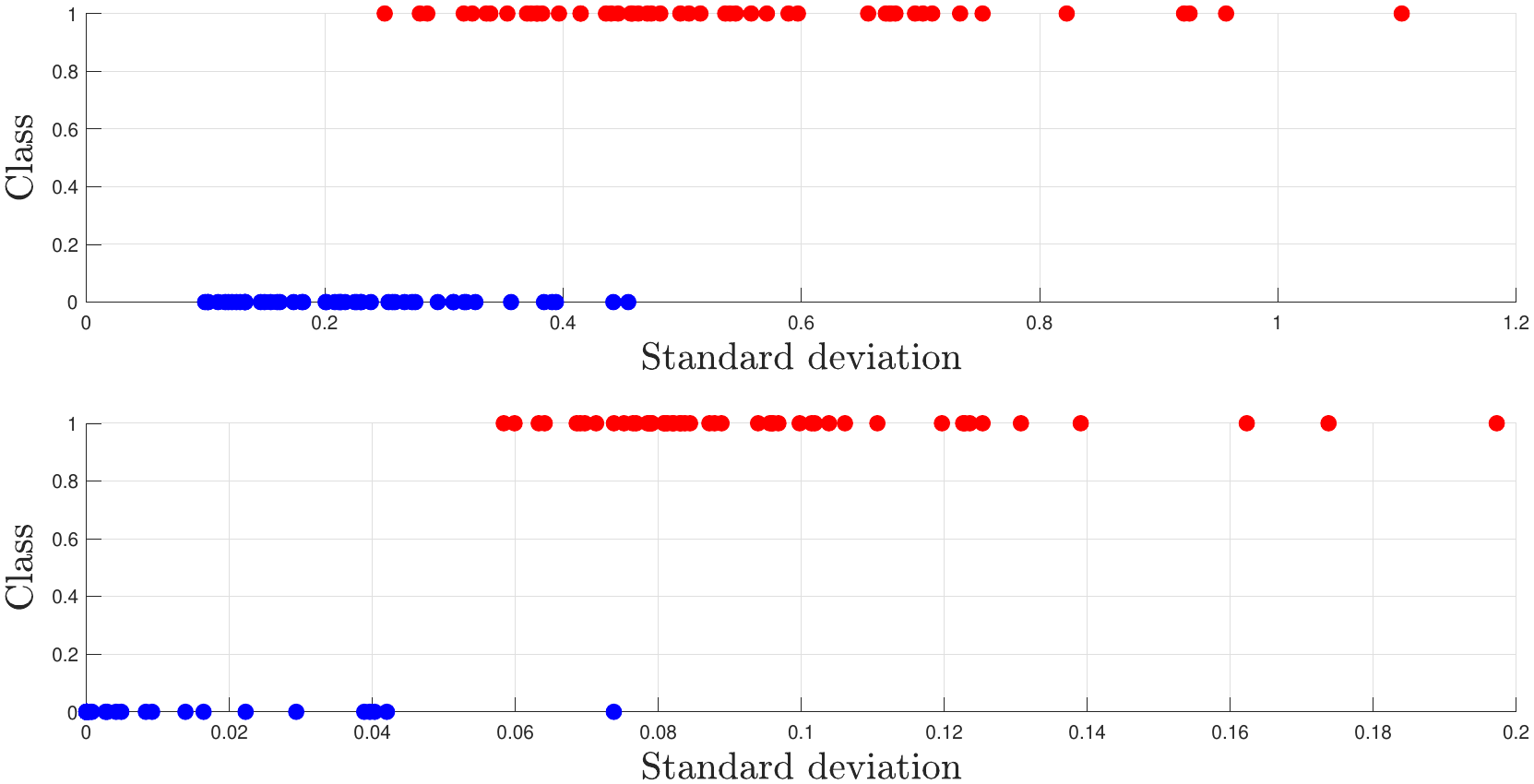}
    \caption{\textbf{TOP:} Standard deviation scores of $100$ randomly selected measurement segments. Red points denote the standard deviation scores of abnormal measurements, while blue points show the scores for measurements acquired on a normal surface. The classes were shifted vertically for better visibility. \textbf{BOTTOM:} Standard deviation scores for the residual signals of the same $100$ measurements.}
    \label{fig:stds}
\end{figure}

\begin{figure}[!t]
    \centering
    \includegraphics[width=0.99\linewidth]{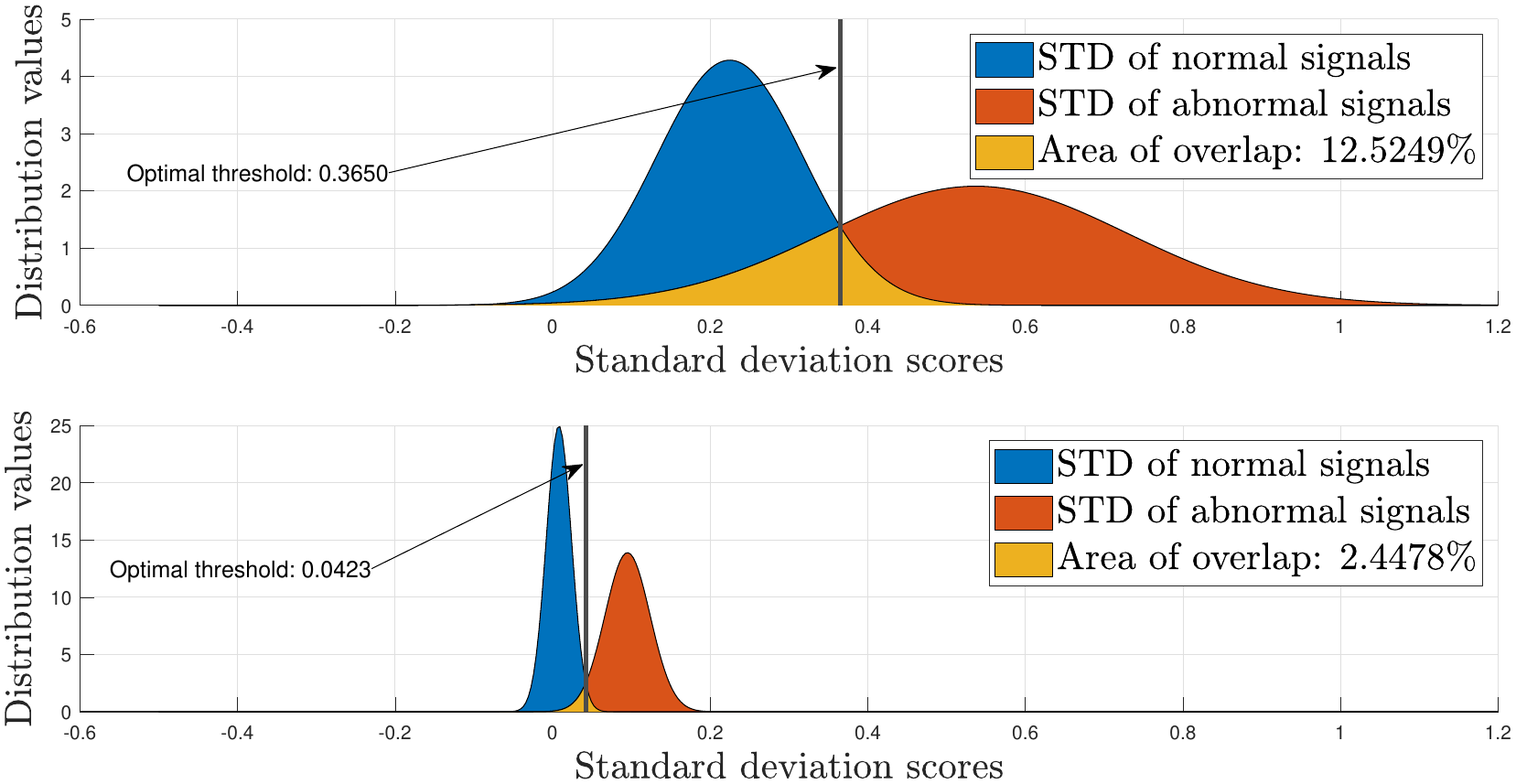}
    
    \caption{\textbf{TOP:} Distributions of standard deviation scores of the measured signals. The vertical line indicates the optimal threshold. \textbf{BOTTOM:} Distributions of standard deviation scores of the transformed signals. The area of overlap is only $\approx 2.5 \%$ of the total area under the distribution curves, whereas it is over $12 \%$ when the proposed transformation is not applied.}
    \label{fig:stddists}
\end{figure}

Our analytic classification approach is based on the assumption that measurements corresponding to abnormal road conditions will contain more noise than segments measured on a normal surface. In order to empirically verify this hypothesis, we randomly selected $100$  force sensor output segments from each class (normal and abnormal) and compared their noise levels. Specifically, we analyzed the standard deviation of each measurement. That is, for a signal segment $\boldsymbol{y} \in \mathbb{R}^{N}$, (see figure \ref{fig:norm_abnorm_periods}), we calculated the corrected sample standard deviation estimate using $N=500$ signal samples as follows:
\begin{equation}
    \label{eq:std_sig}
    s=\sqrt{\frac{1}{N-1} \sum_{k=1}^N | y_k - \mu |^2},
\end{equation}
\noindent where $\mu$ is the mean of $\boldsymbol{y}$. 
In figure \ref{fig:stds}, we computed the standard deviation for the raw signal segments and for the corresponding residual signals given in~\eqref{eq:resid}. It is evident that the proposed residual transformation increased the separability of the two classes. This is also supported by figure~\ref{fig:stddists}, which illustrates the distribution of standard deviation scores belonging to the classes before and after the application of the proposed transformation. Furthermore, let us consider Fisher's discriminant ratio:

\begin{equation}
    \label{eq:fish}
    F_r = \frac{(\mu_n - \mu_a)^2}{\sigma_n^2 + \sigma_a^2},
\end{equation}
where $\mu_n$, $\mu_a$ denote the means, while $\sigma_n$ and $\sigma_a$ denote the variances of standard deviation scores acquired from normal and abnormal signals, respectively. This ratio can be interpreted as a measure for the linear discriminating power of standard deviation scores between the classes. To illustrate this, consider the standard deviation score assigned to each measurement as a projection of that measurement onto the real line. Consider in addition, the means and variances of these projections for each class. If the projected values discriminate well between the classes, then the squared difference of the means is expected to be large, while the within-class variance has to remain small. Calculating~\eqref{eq:fish} for the sensor outputs gives $2.1483$, while the same ratio evaluated over the standard deviation scores acquired from the transformed signals yields $3.6299$. This increase further highlights the effectiveness of the proposed residual transformation in separating the two classes.

Even though the above experiment empirically verified our assumption on the noise levels of the different classes, comparing the standard deviation of residuals does not lead to a perfect classification performance as can be seen in figure \ref{fig:stds}. In other words, the standard deviation of the residual signals may still be a suboptimal feature for indicating road surface abnormalities. To this end, we further improve the classification accuracy by considering data driven classification approaches.

\subsection{A model based data-driven approach}
\label{subsec:vpnet}

In this section we describe the application of VP-NET to the road abnormality detection 
problem. VP-NET is a special neural-network architecture introduced in  \cite{KBHH} containing so-called variable projection layers. These layers (henceforth referred to as VP-layers) are capable of solving \eqref{eq:varpro} and passing the results to a conventional fully-connected neural network.

\begin{figure}[!t]
    \centering
    \includegraphics[width=1\linewidth, trim={0, 7cm, 0, 0}, clip]{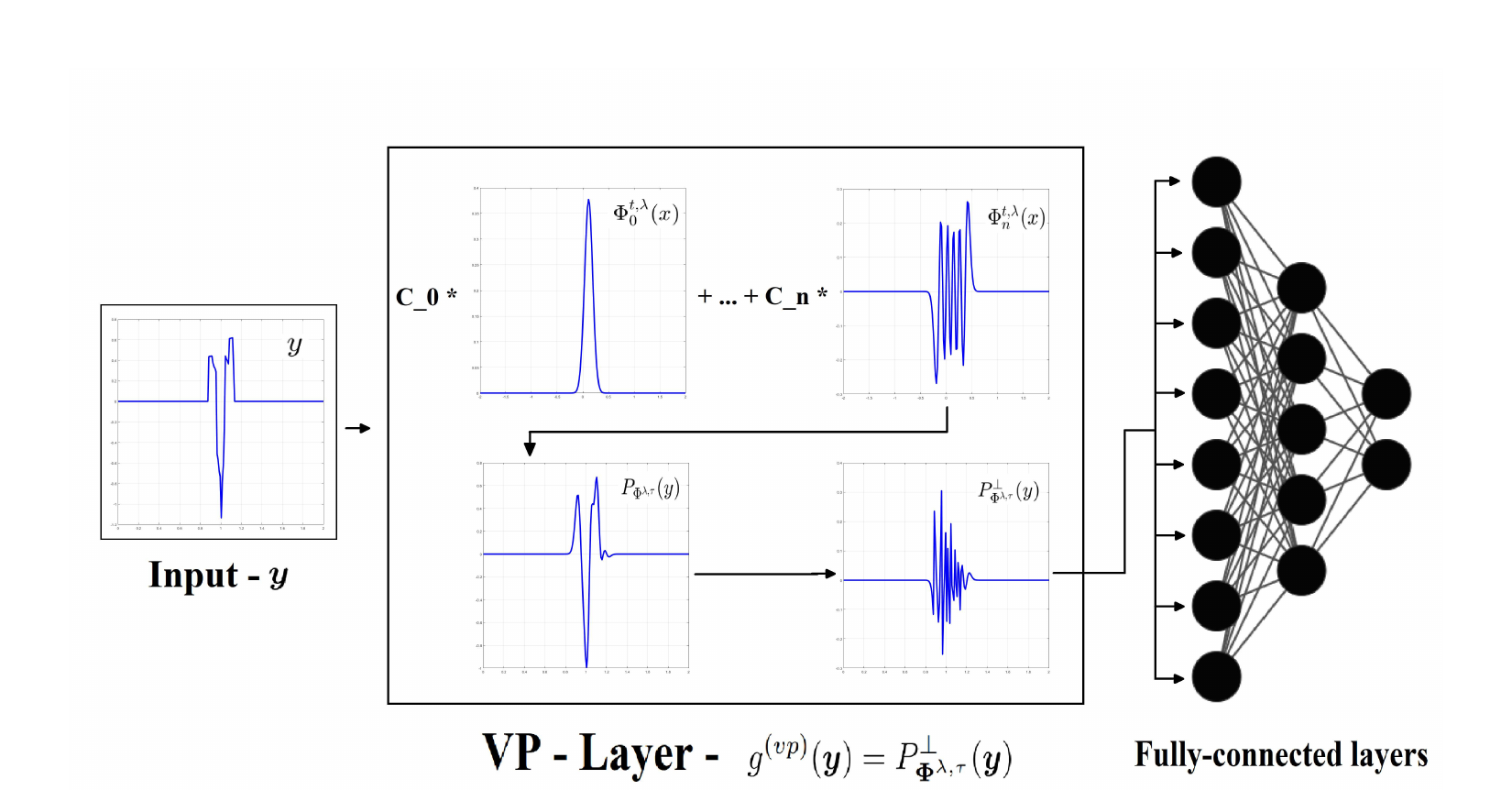}
    \caption{VP-NET architecture. The first layer implements a variable projection operator (in our case an adaptive Hermite function based residual transformation). The output of this layer is then passed to a fully-connected neural network. The parameters of the transformation implemented by the variable projection layer are thus trained together with the weights of the underlying classifier.}
    \label{fig:vpnet}
\end{figure}

The most important benefit of such an approach is that the parameters of the proposed adaptive transformation are trained together with the weights of the classifier. When applied to road surface abnormality detection, VP-NET allows us to simultaneously determine the optimal parameters of the corresponding Hermite function system and the parameters of the classifier that distinguishes between normal and abnormal measurements. VP-layers have several modes of operation. The VP-layer can either pass the linear parameters $\boldsymbol{c}$ in \eqref{eq:linearrep} or the approximaitons~\eqref{eq:proj} and the residuals~\eqref{eq:resid} of the input signal to the subsequent layers. In section \ref{subsec:resid}, we have already established the utility of the adaptive Hermite function based residual transformation~\eqref{eq:resid} to amplify the differences between normal and abnormal measurement segments. Now, we embed this residual transformation into the first layer of the VP-NET architecture via~\eqref{eq:resid}:
\begin{equation}
    \label{eq:vp_layer_res}
    g^{(vp)}(\boldsymbol{y}) = P_{\boldsymbol{\Phi}^{\lambda, \tau}}^{\perp}(\boldsymbol{y})\qquad (\boldsymbol{y}\in\mathcal{F}).
\end{equation}
The output of the first VP-layer $g^{(vp)}(\boldsymbol{y})$, i.e., the residual signal, is forwarded to the subsequent fully-connected layers of the network, which are responsible for classification. A visual representation of the VP-NET architecture used in this study is given in figure \ref{fig:vpnet}.

Convolutional layers are also commonly used to learn relevant representation of the data, however as can be seen in section \ref{sec:tests}, the application of VP-NET is more appropriate for detecting road surface abnormalities. The reasons behind this could be traced back to several important differences between VP-NET and convolutional neural networks. Instead of learning the appropriate kernel weights (which could be numerous), the VP-layer optimizes only a few nonlinear parameters of the underlying adaptive transformation. In our case for example, the parameter vector to be learned contains only two components: $(\lambda,\tau) \in \mathbb{R}^2$, i.e., the dilation and the translation parameters of the adaptive Hermite functions. In contrast, the best performing convolutional neural network solving the abnormal surface detection problem used a kernel size of $25$ increasing the complexity of the network architecture, and thus the computational cost of the training process. Additionally, the parameters of the VP-layer are interpretable representing the location and the width of the spike-like waveforms in the input~\cite{KBHH}, whereas the weights of the convolutional layers have no such physical meaning.

\section{Classifier specifications}
\label{sec:classSpecs}

In total, we compared the effectiveness of $6$ classification schemes. The examined methods included analytical (noise thresholding of the raw and the transformed measurements) approaches, as well as classical data-driven (SVM, neural networks) and model-driven (VP-NET) machine learning algorithms. When applicable, the hyperparameters of the classifier (e.g. number of layers, etc.) were identified using an exhaustive grid search of the parameter space. 

In section \ref{subsec:resid}, we found that abnormal signal segments are expected to have higher standard deviation score compared to normal segments. Using this observation, we introduced a classification approach based on simple thresholding. First, we split the available measurements into training and test sets. Then, a standard deviation threshold was identified which maximizes the classification accuracy on the training set. This was done iteratively starting with the lowest standard deviation score in the training set, which was increased using a small step size until the optimal classification accuracy was achieved. Once the threshold value is fixed, a single measurement can be classified as "abnormal" if its standard deviation falls above this optimized threshold. We used the same thresholding algorithm for classifying the transformed measurements as well. That is, we first applied the Hermite based residual transformation~\eqref{eq:resid} to every available measurement segment, then calculated an optimized noise threshold for the transformed training set. Since the residual transformation removes low frequency components from the measurements while leaving the noise levels intact, we expect the second approach to provide significantly better accuracy scores. Even though the discussed analytic methods proved to be somewhat less accurate than their data driven counterparts (see Table~\ref{tab:scores}), their appeal is in their simplicity. Once an optimal threshold has been identified based on the training data (offline), one can easily provide an implementation of this classification scheme capable of running on the limited hardware resources available in a commercial vehicle. Implementing the proposed transformation and the trhesholding based abnormality detection scheme on microcontrollers or FPGAs (Field Programming Gate Arrays) is conceivable.

Now, we specify the model-based data-driven classifier discussed in \ref{subsec:vpnet}. As illustrated in Figure \ref{fig:vpnet}, VP-NET is made up of two parts: VP-layers are responsible for automatic feature extraction, and fully-connected layers are responsible for classification. After hyperparameter optimization via grid search, we settled on using a VP-NET architecture consisting of a single VP-layer followed by $3$ fully-connected layers. The fully-connected layers each contained $64$ neurons with ReLu activation. The VP-layer implemented the residual transformation \eqref{eq:vp_layer_res}, using $n=11$ adaptive Hermite functions. Note that the resulting VP-NET architecture is small enough to be used with limited hardware resources, such as microcontrollers, for future online road surface abnormality detection tests.

In addition to the proposed methods discussed above, we also included some well-known data driven classification schemes in our experiments. This allowed us to compare the effectiveness of the proposed approaches to classical machine learning algorithms. The classifiers in question included a support vector machine (SVM) with linear kernel, a fully-connected neural network (FCNN), and a convolutional neural network (CNN). The inputs of each algorithm were the preprocessed force sensor measurement segments (as discussed in section~\ref{sec:preproc}), i.e., the residual transformation was excluded from preprocessing pipeline. This way, the effectiveness of the proposed VP-NET architecture can be compared to those learning approaches which operate on the raw data and extracts features automatically. To this end, we included a convolutional neural network consisting of a single convolution layer with a single filter and a kernel size of $25$. The architecture resembles the VP-NET in the sense that the CNN also provides an automatic feature extraction step via the first layer followed by fully-connected layers which are responsible for the classification. For better comparability, the simple FCNN as well as the fully-connected part of the CNN were matched to the fully-connected part of the proposed VP-NET architecture, while the number of filters and the corresponding kernel sizes of the CNN were optimized in an exhaustive manner using a grid search of the parameter space. 

\section{Experiments}
\label{sec:tests}

\begin{figure}[!htb]
    \centering
    \includegraphics[width=1\linewidth, trim={35mm, 0, 0, 20cm}, clip]{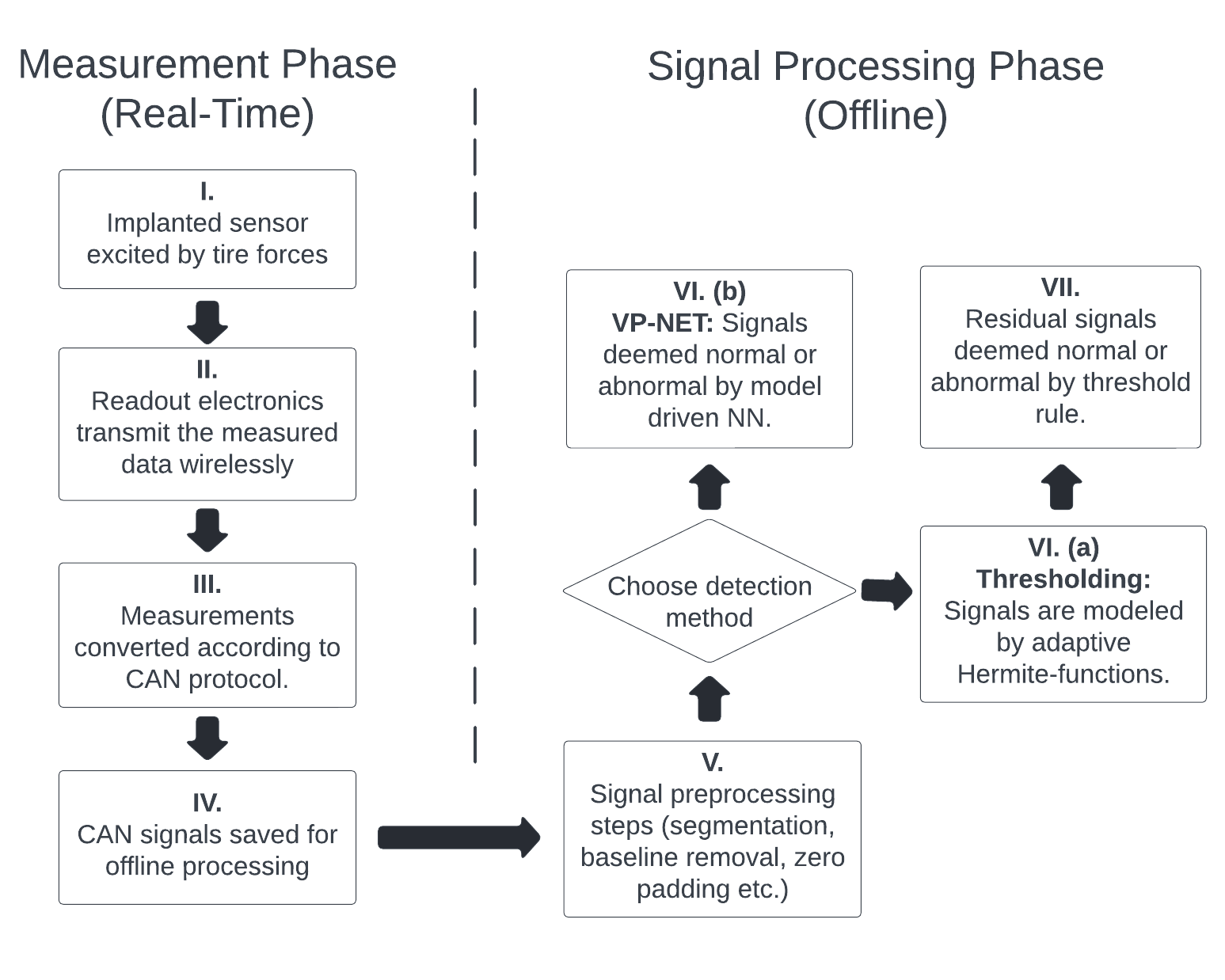}
    \caption{A schematic representation of the proposed method. Measurements were observed online, whereas the abnormality detection was performed offline on the previously measured signals.}
    \label{fig:schematic}
\end{figure}

In this section, we evaluate the performance of the proposed analytic and data-driven surface abnormality detection methods. A block diagram of the proposed measurement system is provided in Fig. \ref{fig:schematic}. The experiments for this study have been conducted in a two phase, offline and online manner. First, we used our test vehicle setup described in section~\ref{sec:exper} to gather measurements in real-time. Then, the force sensor output signals were saved for later processing. The signal processing part of our measurement system is responsible for applying the proposed orthogonal transformations to the sensor outputs (modeling) and for identifying abnormal tire revolutions (classification). Regardless of whether we choose the VP-Net approach described in \ref{subsec:vpnet} or the threshold based method from \ref{subsec:resid}, a parameter optimization phase known as training is required for the signal processing algorithms. Training involves finding the optimal values of the model parameters, which maximize the precision of the abnormality detection algorithms on a set of already labeled measurements referred to as the training set. The training process should be performed offline after a sufficient amount of data has been collected. This phase is of utmost importance, since the choice of the training set influences the generalization power of the classifiers. For instance, training with imballanced datasets in which the "normal" tire revolutions are over represented, can reduce the ability of the classifier to recognize "abnormal" signals. Once the parameters have been optimized, the classifier can be used to detect abnormal measurements which were not included in the training set. This evaluation phase has also been done offline in our study, but the simplicity of the proposed algorithms enables real-time abnormality detection, which will be a part of our future work. The training of our algorithms were carried out in a supervised framework, which requires ground truth data. In other words, we have to label the gathered measurements as "normal" or "abnormal" before the training and the testing procedures take place. Since our measurements were taken on the public roads of Budapest without the ability to pinpoint the exact occurrence of road surface abnormalities, in this paper we relied on automatic labeling schemes. 

\begin{figure}[!htb]
    \centering
    \includegraphics[width=1\linewidth]{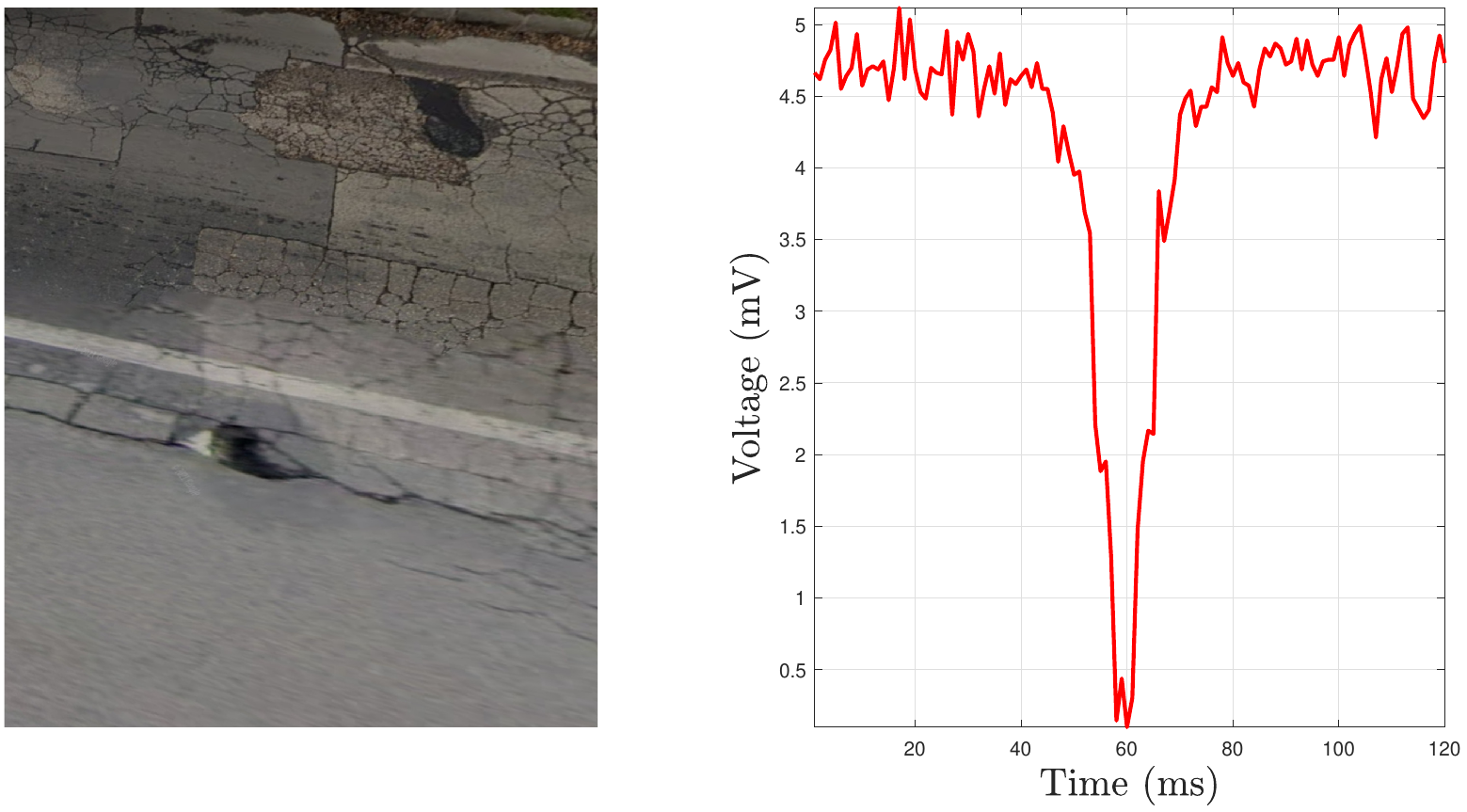}
    \caption{\textbf{LEFT:} Road abnormalities (in this case potholes) from our measurement route. Image taken from Google Maps (coordinates 47.52, 19.08). \textbf{RIGHT:} Wheel sensor output taken from the same measurement, classified as abnormal.}
    \label{fig:vaganyutca}
\end{figure}

Recently many such algorithms have been proposed \cite{CLTW, HGV, WSDM, DQGH}, which rely either solely on acceleration data (along the $X$ and $Z$ axes), or on various sensor fusion strategies. The former approach is often based on some variation of the so-called Gaussian background model which we also utilized in this study. Actually, the vertical acceleration data from the accelerometer attached to the chassis of our test vehicle is used for automatic ground truth generation according to Ref.~\cite{DQGH}. For the "abnormal" tests, our vehicle passed through poor quality asphalt with potholes and manhole covers. In the following results "abnormal" data samples will refer to tire revolutions where the vehicle encountered these obstructions. Figure~\ref{fig:vaganyutca} illustrates an abnormal road surface on our test route and a wheel sensor output segment from the same measurement.

In~\cite{DQGH}, the authors show that their proposed labelling scheme is capable of detecting potholes and manhole covers, making their approach suitable for the labelling of our measurements. Our results thus should be viewed as proof of concept, showing that the proposed measurement system does indeed provide similar results to the state-of-the-art. As part of our future work, we plan to record measurements, where the exact locations of surface abnormalities are known in advance. This would allow to implement more objective test scenarios. There are reasons to be optimistic about such comparisons however, as the proposed measurement system would be expected to be more sensitive than said acceleration based approaches. This is because instead of measuring changes in acceleration at a single point of the vehicle, our approach relies on the changes in the dynamics at each wheel, thus providing a finer resolution of the surface.

In order to ensure a fairly balanced dataset, the measurements used for our experiments contained roughly the same amount of normal and abnormal signal segments (see Table~\ref{tab:data_desc}). This setup also makes classification accuracy~\eqref{eq:acc} suitable for evaluating the performance of the proposed classifier models. 
\begin{table}[!t]
\caption{Number of signal segments in each class.}
\begin{center}
\begin{tabular}{|c | c | c|} 
 \hline
 \textbf{Normal signal segments} & \textbf{Abnormal signal segments} & \textbf{Total}\\ [0.5ex] 
 \hline
 $282 \ (54.55\%)$ & $235 \ (45.45\%)$ & $517$ \\ 
 \hline
\end{tabular}
\end{center}
\label{tab:data_desc}
\end{table}
The algorithms specified in section \ref{sec:classSpecs} were evaluated using $5$-fold cross validation.  At each fold, the training set contained $413 \ (80\%)$ measurement segments, with the remaining $104 \ (20\%)$ signals assigned to the test set. Note that our training and test sets were disjoint, and no information was shared between the training and the testing phases. Table \ref{tab:scores} shows the average accuracy achieved on the test set across every fold, as well as the lowest and highest accuracy scores for a single fold. The classifiers "Threshold" and "Threshold Hermite" refer to the standard deviation thresholding methods we proposed in section~\ref{subsec:resid}.

\begin{table}[!t]
    \caption{Examined classification schemes and their accuracy.}
    \centering
    \begin{tabular}{|c|c|c|c|}
    \hline
        \textbf{Classifier} & \textbf{Mean Accuracy} & \textbf{Highest} & \textbf{Lowest} \\
    \hline \hline
        Threshold & $86.92 \%$ & $90.38 \%$ & $79.81 \%$ \\
    \hline
        SVM & $93.65 \%$ & $96.15 \%$ & $92.31 \%$\\
    \hline
        Threshold Hermite  & $93.85 \%$ & $96.15 \%$ & $90.38 \%$ \\
    \hline
        FCNN & $96.13 \%$ & $99.02 \%$ & $94.23 \%$ \\
    \hline
        CNN & $96.52 \%$ & $97.12 \%$ & $93.20 \%$ \\
    \hline
        VP-NET & $\boldsymbol{97.68 \%}$ & $\boldsymbol{99.04 \%}$ & $\boldsymbol{96.12 \%}$ \\
    \hline
    \end{tabular}
    \vspace{1mm}
    \label{tab:scores}
    \vspace{-0.5cm}
\end{table}

Our results provide a solid proof of concept for the efficiency of the proposed measurement system in detecting road surface abnormalities. Observe that the low-dimensional Hermite-representation and the corresponding residual transformation significantly improved the detection rates. Indeed, when considering analytic approaches, the increase in accuracy is particularly high (${\approx}7 \%$). Actually, the simple "Thershold Hermite" approach even outperforms the much more sophisticated SVM method, where no residual transformation is applied. The results are also promising for the examined data-driven classifiers. Here, the algorithm incorporating the proposed residual transformation (VP-NET) again provides the best accuracy score. Although the performance increase this time is not as high, the VP-NET architecture is significantly simpler than the second best CNN architecture. In this particular case, the training of the VP-layer involved the optimization of $2$ parameters, while $25$ kernel weights had to be optimized for the CNN.  Overall, we acquired promising results supporting our claim that the wheel sensor and the proposed classifier models can be efficiently used for surface abnormality detection.

\section{Conclusion and future plans}
\label{sec:conclusion}

In this paper we proposed a measurement system
for intelligent tires that is based on a 3-dimensional piezoresistive force sensor. To demonstrate its efficiency, the problem of road surface abnormality detection was chosen. We developed analytic model-based as well as data-driven machine learning approaches to process and classify the output signals produced by the sensor. The algorithms were evaluated in a real-world test scenario, and we showed that both classification approaches (analytic and data-driven) produce reliable results ($93 \%$ to $97 \%$ detection accuracy).  It also turned out that relevant features can be extracted via Hermite-representation and residual transformation, which improves the accuracy of basic classifier models. 

Considering the simplicity of the proposed classifiers, especially the analytic approaches, running the pre-trained algorithms on microcontrollers or FPGAs should pose little difficulty. This allows for online abnormality detection that can be utilized in other autonomous vehicle applications, such as road quality estimation, and autonomous speed control, which will be part of our future work.

\section*{Acknowledgments}
This project was supported by the J\'anos Bolyai Research Scholarship of the Hungarian Academy of Sciences. Project no.~TKP2021-NVA-29 has been implemented with the support provided by the Ministry of Innovation and Technology of Hungary from the National Research, Development and Innovation Fund, financed under the TKP2021-NVA funding scheme. This work was supported by the National Research, Development and Innovation Fund of the Hungarian Government in the framework of KoFAH, NVKP 16-1-2016-0018. The authors thank the contribution of Attila Nagy for sensor integration and the valuable work of Dr. Péter Földesy in the design and implementation of the read-out electronics. This project was supported by the NVKDP Cooperative Doctoral Program by the Hungarian Ministry of National Development and the National Research, Development and Innovation Fund.  The research was supported by the Ministry of Innovation and Technology NRDI Office within the framework of the Autonomous Systems National Laboratory Program.

\bibliographystyle{IEEEtran}
\bibliography{pub}

\end{document}